\documentclass[
prl,
twocolumn,
nofootinbib,
tightenlines %
floatfix,
showpacs ,amssymb, aps,
superscriptaddress
]{revtex4-1}
\usepackage{amsmath,slashed}
\usepackage[left]{lineno}

\usepackage{comment}
\usepackage{graphicx}
\usepackage{amsmath,amssymb}
\usepackage{bm}
\usepackage[shortlabels]{enumitem}
\usepackage{float,soul}
\usepackage{natbib}
\usepackage{wasysym}
\usepackage{aas_macros}
\graphicspath{{images/}}
\RequirePackage{lineno}
\usepackage[usenames,dvipsnames]{color}

\newcommand{\SPA}{School of Physics and Astronomy, Monash University, Vic 3800, Australia}
\newcommand{\OzGravMonash}{OzGrav: The ARC Centre of Excellence for Gravitational Wave Discovery, Clayton VIC 3800, Australia}

\begin{document}

\title{Computing Fast and Reliable Gravitational Waveforms of Binary Neutron Star Merger Remnants}

\author{Paul J. Easter}
\email{paul.easter@monash.edu}
\author{Paul D. Lasky}
\email{paul.lasky@monash.edu}
\affiliation{\SPA}
\affiliation{\OzGravMonash}
\author{Andrew R. Casey}
\email{andrew.casey@monash.edu}
\affiliation{\SPA}
\affiliation{\OzGravMonash}
\affiliation{Faculty of Information Technology, Monash University, VIC 3800, Australia}
\author{Luciano Rezzolla}
\affiliation{Institut f\"ur Theoretische Physik, Max-von-Laue-Stra\ss e 1, 60438 Frankfurt, Germany}
\author{Kentaro Takami}
\affiliation{Kobe City College of Technology, 651-2194 Kobe, Japan}
\pacs{
}

\begin{abstract}
	Gravitational waves have been detected from the inspiral of a binary neutron-star, GW170817, which allowed constraints to be placed on the neutron star equation of state. The equation of state can be further constrained if gravitational waves from a post-merger remnant are detected. Post-merger waveforms are currently generated by numerical-relativity simulations, which are computationally expensive. Here we introduce a hierarchical model trained on numerical-relativity simulations, which can generate reliable post-merger spectra in a fraction of a second. Our spectra have mean fitting factors of 0.95, which compares to fitting factors of 0.76 and 0.85 between different numerical-relativity codes that simulate the same physical system. This method is the first step towards generating large template banks of spectra for use in post-merger detection and parameter estimation.
\end{abstract}
\maketitle
    Gravitational waves have been  observed from the inspiral of binary neutron star merger GW170817~\cite{GW170817Detection}. This allowed limits to be placed on the neutron star tidal deformability ~(see~e.g.,~\cite{GW170817Detection,Annala2018,Radice2018,Most2018,De2018,GW170817Properties}). However, due to lack of detector sensitivity at high frequencies, the merger and post-merger signals were not detected~\cite{GW170817Properties,GW170817postmerger1,GW170817Postmerger2}. The probability of post-merger signal detection increases as the high-frequency sensitivity of gravitational-wave interferometers  improves.   \par
    The detection and characterisation of a post-merger remnant is aided by a large bank of gravitational-wave strain waveforms. Generating such waveforms is currently computationally expensive, and there are only $\mathcal{O}(100)$  in existence. In this work we make a step towards generating a large template bank of post-merger spectra by training a hierarchical model on a set of numerical-relativity spectra.\par    
    There has been significant research applied to the relationship between post-merger numerical-relativity simulations, the corresponding spectrum of the gravitational-wave strain, and the neutron star equation of state~(e.g.,~\cite{Stergioulas2011,Bauswein2012,read13,hotokezaka13,Takami2014,bernuzzi2015,Takami2015,Clark2016postmerger,rezzolla16,Radice2017,Dietrich2017,Radice2017a,Pietri2018,Dietrich2018,Dietrich2018a,Bose2018}). There are many degrees of freedom for each simulation, which include the neutron star system parameters, equation of state, and  simulation parameters, as well as parameters related to magnetic fields and neutrinos. We choose to use a set of numerical-relativity simulations that are homogeneous, eliminating unwanted variations between simulations with different parameters. To achieve this, we use a subset of 35 waveforms from \citet{rezzolla16} consisting of identical simulation parameters with variations in the neutron star mass and equation of state only. To obtain consistent spectra, we select waveforms that have a fixed time-step and almost identical length. \par
    \citet{Clark2016postmerger} showed that dimensional reduction of post-merger waveforms is possible by performing principal component analysis after aligning the maximum of each gravitational-wave strain spectra in the frequency domain (see also \cite{Bose2018}). We use a similar method of frequency shifting in our model. We introduce a hierarchical model that trains on existing numerical-relativity post-merger simulations, and can produce new, accurate spectra in a fraction of a second. This is the first step towards making large template banks of post-merger spectra suitable for detection and parameter estimation which could compliment existing unmodelled searches for post-merger remnants~\cite{GW170817Properties,Klimenko2016,Chatziioannou2017}. \par
    Simulation of the post-merger phase of binary neutron star mergers is significantly more complicated than the inspiral phase due to the complex physics including shock heating and nonlinear mode coupling. Additional effects, such as neutrino cooling and magnetic fields, are not expected to yield substantial modifications to the locations of the spectral peaks (see e.g. \cite{Giacomazzo2011,Sekiguchi2016,Kawamura2016}), while the role of viscous effects is still a matter of debate \cite{alford2018}. The accuracy of the resulting simulations can be limited by the trade off between computational constraints and higher resolutions~\cite{Baiotti2017}. This is particularly true for the phase evolution of the post-merger simulations which do not necessarily converge~\cite{Dietrich2018}. However, the power-spectral content is convergent for sufficiently high resolutions (e.g.,~\cite{Takami2015,Dietrich2018}). Our model is representative up to the validity of the numerical-relativity simulations that it is based upon. With this in mind, we wish to encourage further research into numerical-relativity simulations of post-merger remnants to increase the available number of waveforms and to enable further cross-checking between codes.\par
\begin{figure*}
        \includegraphics[width=0.94\textwidth]{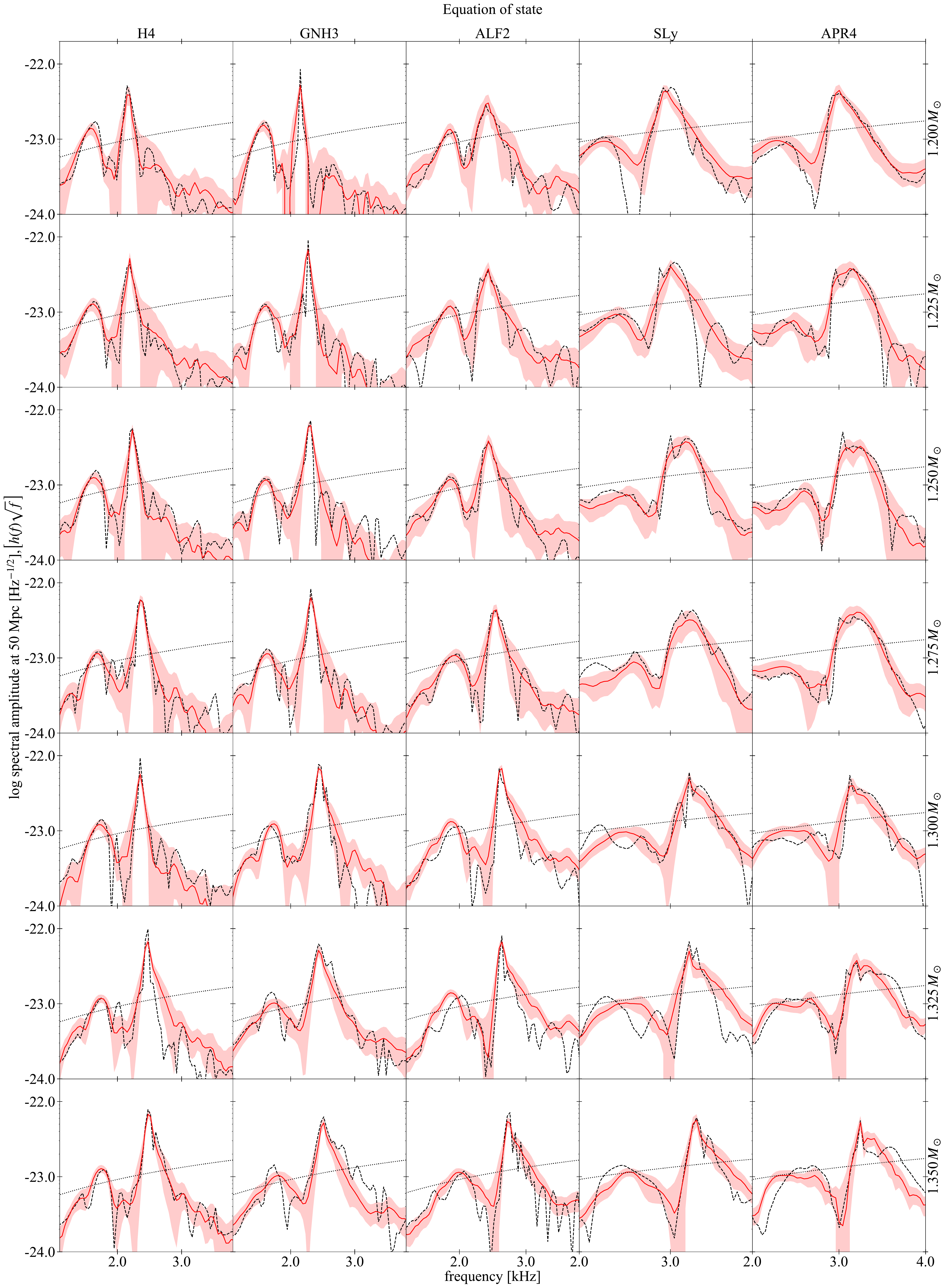} 
        \caption{ Reconstructed gravitational-wave spectra generated with leave-one-out cross-validation (solid red) and original numerical-relativity spectra (dashed black), scaled to a distance of 50 Mpc. Each column represents a different equation of state and each row represents a different neutron star mass, increasing towards the bottom. The one-sigma uncertainty in the spectra is shaded in light red for each prediction.  The Advanced LIGO noise amplitude spectral density (dotted black curve)~\cite{PSD:aLIGO} is shown on all subplots. }
        \label{fig:GoodPlotsWithWaveformGeneration}
\end{figure*} 
        We use 35 numerical-relativity simulations of binary neutron star mergers from~\citet{rezzolla16}, to which we refer to for details on the equations of state employed.  Each simulation consists of non-spinning, equal mass progenitor neutron stars, with five different equations of state across the simulations.  We train our model on the amplitude  of the characteristic strain spectra, $h_c(f)~=~|\tilde{h}(f)|\sqrt{f}$. Here,  $\tilde{h}(f)$ is the Fourier transform of the plus polarisation of the post-merger gravitational-wave strain, $h_+(t>0)$. The plus and cross polarisations of the simulated gravitational-wave strain have almost identical spectral amplitude and have a phase offset of almost exactly $\pi/2$. We gain no extra information by including the cross polarisation. The merger time, $t=0$, is defined as the time where $h_+^2(t)~+~h_\times^2(t)$ reaches the first maximum. \par
        
        We use a hierarchical model to represent the amplitude spectra. Given a neutron star of mass $M$ and radius $R$, we assume the compactness of a neutron star, $C\equiv M/R$, in the $j$th simulation has a power-law dependence with the mass, $M$, over all equations of state
    \begin{align}
        C_j = \alpha_j M_j^{\beta_j}. 
        \label{eq:Ccalc}
    \end{align}
        The validity of this model will be determined by how well we can match the numerical-relativity waveforms. The hyperparameters $\{\mathbf{a},\mathbf{b}\}$, and the quadrupolar tidal deformability, $\kappa_2^\tau$, determine the values of $\{\alpha,\beta\}$ as follows:
    \begin{align}
        \alpha_j & \sim\mathcal{N}(a_0 + a_1 \kappa_{2,j}^\tau,\sigma_\alpha^2),\label{eq:alphacalc} \\ 
        \beta_j & \sim\mathcal{N}(b_0 + b_1 \kappa_{2,j}^\tau,\sigma_\beta^2),
        \label{eq:betacalc}
    \end{align}
        where $\mathcal{N}(\mu,\sigma^2)$ is a Gaussian distribution of mean $\mu$ and variance $\sigma^2$. The quadrupolar tidal deformability, $\kappa_2^\tau$, is used due to its importance in the inspiral dynamics~\cite{Damour2010,read13,Takami2014,bernuzzi2015,Takami2015,rezzolla16} and its correlation with the location of the main frequency peak of the post-merger spectrum~\cite{Bauswein2012,Takami2014,bernuzzi2015, Takami2015}.\par
        All spectra in the training set, which excludes the spectrum under test when leave-one-out cross-validation is performed, are used to determine the hyperparameters $\{\mathbf{a},\mathbf{b}\}$ by a least squares fit. The amplitudes for each spectra are frequency shifted so that the peak frequencies are aligned in a similar way to Refs. \cite{Clark2016postmerger,Bose2018}.  We then fit the aligned spectral amplitudes with a linear model
    \begin{align}
        (h_c)_{i,j} = \boldsymbol{\Theta}_{i}\mathbf{X}_j + \text{noise}, 
        \label{eq:GenerateWaveform}
    \end{align}
        where the noise is modelled as intrinsic variance, $s_i^2$ for the $i$th frequency bin, $\boldsymbol{\Theta}_i$ is a vector of unknown coefficients,  and $\mathbf{X}_j$ is a design matrix of
    \begin{align}
        \mathbf{X}_j=[1,\widehat{C}_j,\widehat{M}_j,\widehat{\kappa^\tau_{2,}}_{j}].
    \end{align}
       \noindent The hats indicate the whitened transformations of the neutron star parameters such that $\widehat{\mathbf{x}}\sim~ \mathcal{N}(0,1)=( \mathbf{x}-\mu_x)/\sigma_x$ where $\mu_x$ and $\sigma_x^2$ are the mean and variance of $\mathbf{x}$ respectively. Spectra can then be trivially generated given any mass, quadrupolar tidal deformability and frequency shift. The frequency shift can be determined from the value of the quadrupolar tidal deformability~\cite{bernuzzi2015,Takami2015}. \par
       We perform leave-one-out cross-validation to test the performance of the model. We do this by excluding the spectrum under test and its associated parameters from the training set. In doing so, the spectra generated during leave-one-out cross-validation represent an extrapolation by the model  and the fitting factors are therefore conservative.
       \par
       We perform spectral comparisons using the following noise-weighted fitting factor, or overlap~\cite{apostolatos95}
        \begin{equation}
        	FF(h_{1},h_{2})\equiv\frac{\left<h_{1}|h_{2}\right>}{\sqrt{\left<h_{1}|h_{1}\right>\left<h_{2}|h_{2}\right>}}.\label{eq:FF}
        \end{equation}
        Here, the inner product is defined by
        \begin{align}
        	\left<h_1|h_2\right>\equiv4 \int df\frac{|\tilde{h}_1(f)|\ |\tilde{h}_2(f)|}{S_h(f)},
        \end{align}
        where $S_h$ is the noise power spectral  density. The resultant fitting factor is similar to the standard fitting factor except that it operates on the Fourier amplitude only. A fitting factor of one represents a perfect match.\par
        While it is known that smooth relationships exist between various system properties (e.g., mass, tidal parameters, etc.) and post-merger waveform spectral features~\cite{Bauswein2012,Takami2014,bernuzzi2015,Takami2015,Bauswein2015,rezzolla16}, no such relationships exist for the phase information~(see also~\cite{messenger14}). Empirically, while we find good training fits using our model on the spectral content of the waveforms (see below), we are not able to train successfully or robustly on the full time series including both phase and amplitude as the phase evolves too quickly between adjacent simulations.  We discuss this in more detail below. \par 
    \begin{figure}[H]
          \includegraphics[width=0.99\columnwidth]{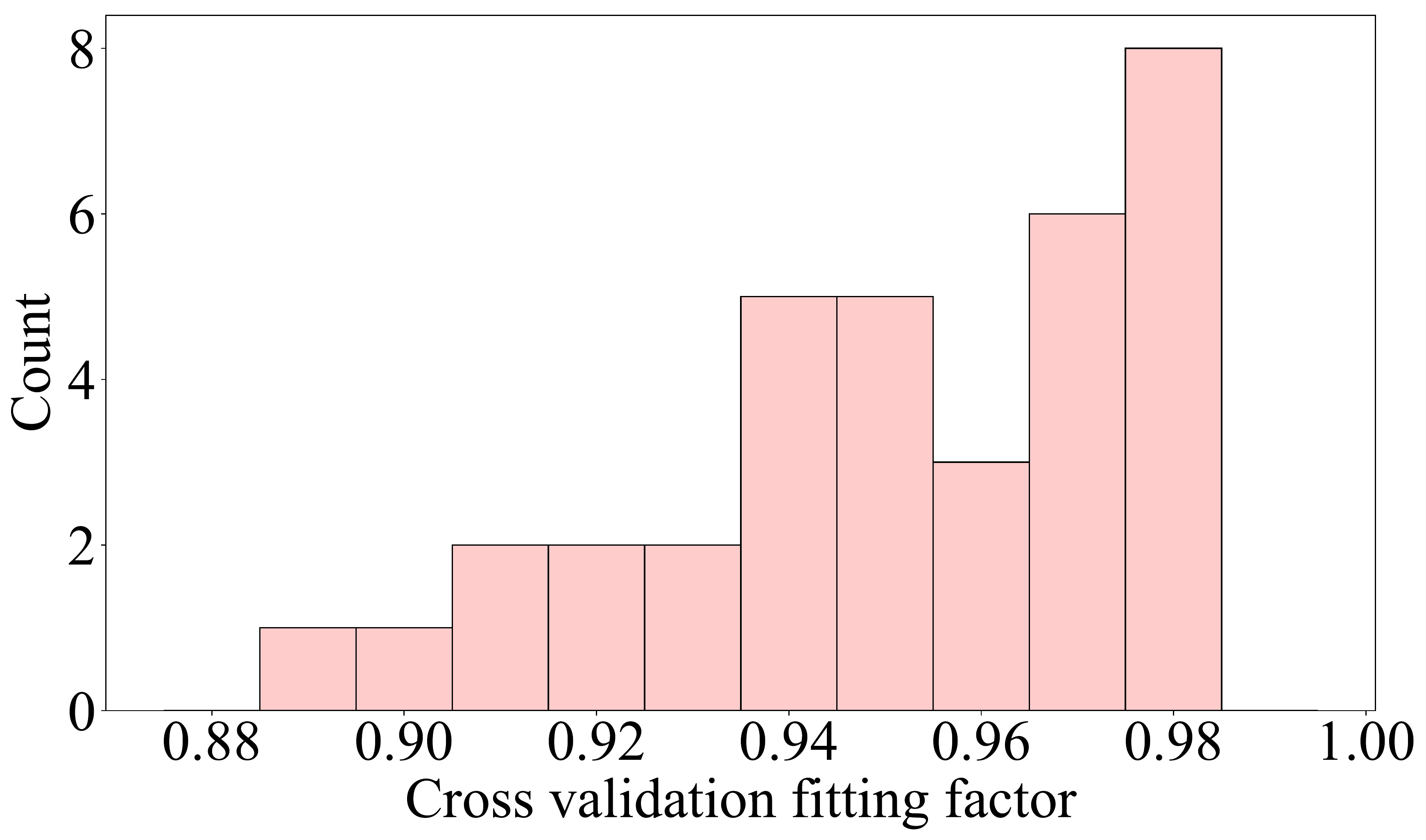}  
          \caption{Histogram of fitting factors determined by comparing numerical-relativity spectra with spectra generated by our model using leave-one-out cross-validation}.
          \label{fig:FittingFactorsHistogram}
    \end{figure}
    \begin{figure*}
        \includegraphics[width=\textwidth]{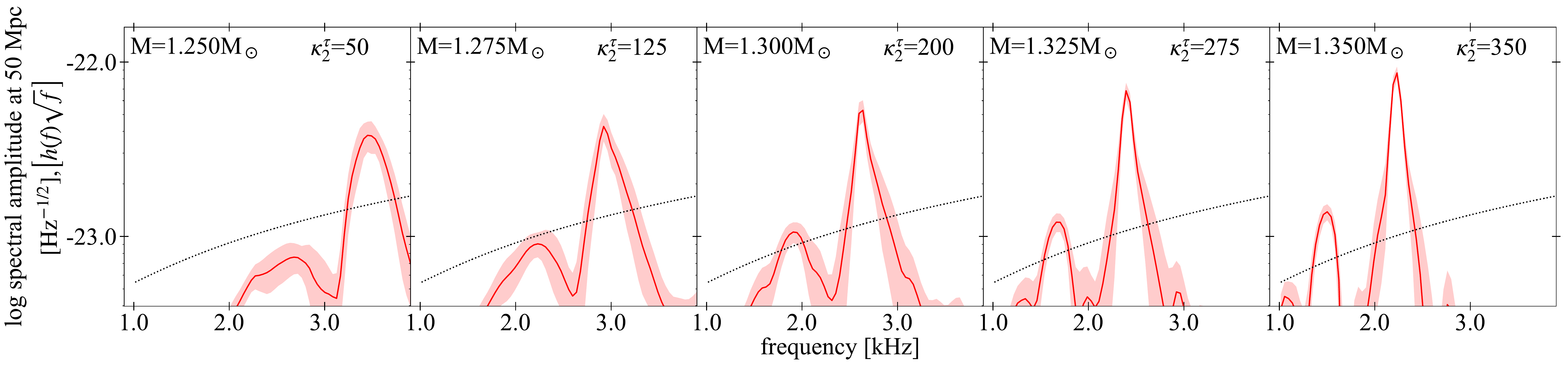} 
        \caption{ Spectra generated by the model when trained on all the numerical-relativity spectra (red). The uncertainties in the spectra are shown in light red. The Advanced LIGO noise amplitude spectral density (dotted black) is shown on all subplots.}
        \label{fig:GeneratedWaveforms}
    \end{figure*}    
        Following the training step, we use Eq.~(\ref{eq:GenerateWaveform}) to generate spectra.  Figure~\ref{fig:GoodPlotsWithWaveformGeneration} shows how well our generated spectra match the original numerical-relativity spectra. The original spectra are shown as dashed black curves, the cross-validation spectra are shown as red curves, and the one-sigma model uncertainty is shown in shaded light red. All spectra are scaled to a distance of 50 Mpc. The Advanced LIGO noise amplitude spectral density is shown as the dotted black curves~\cite{PSD:aLIGO}.  We fit the large-scale structure of the numerical-relativity peaks well with some deviations in the small-scale structure. \par Figure~\ref{fig:FittingFactorsHistogram} shows a histogram of the noise-weighted fitting factor, Eq.~(\ref{eq:FF}), between our cross-validated model prediction and the corresponding numerical-relativity spectra for all tested waveforms. The resulting histogram has a mean of 0.95 with a standard deviation of 0.03. This indicates that the large-scale fit is more important than any small-scale deviations.\par
        As a baseline value for comparison, we compare fitting factors between the numerical-relativity spectra used in this paper, and those produced with other codes~\cite{Dietrich2018}. Notwithstanding the fact that post-merger waveforms can differ with resolution even when using the same code, we assume that the waveforms have similar truncation errors and compare two such sets of spectra using equations of state H4~\cite{bernuzzi2015a} and SLy~\cite{Dietrich2017a} for equal mass binaries with $M=1.35\,M_\odot$. 
        The positions of the largest-amplitude spectral peaks and their uncertainties are indeed comparable between different codes, but
        the fitting factor for each set of waveforms is 0.76 and 0.85 for H4 and SLy, respectively.  This indicates that our fitting factors are better than fitting factors between different numerical-relativity codes. We discuss the implications of this below.\par
        To evaluate how the generated spectra vary, we train on all numerical-relativity spectra and generate a grid of model spectra. We generate spectra at five equally spaced mass and quadrupolar tidal deformability values. The mass ranges from $M=1.25\,M_\odot$ to 1.35\,$M_\odot$, and  the quadrupolar tidal deformability varies from $\kappa_2^\tau = 50$ to 350. The generated spectra are shown in Fig.~\ref{fig:GeneratedWaveforms} as the red curves,  the one-sigma model uncertainty as light red shading, and the Advanced LIGO noise curve as dotted black. Each of these spectra take a fraction of a second to evaluate. We show these spectra to indicate what we can hope to achieve by implementing these models in full parameter estimation. \par
        
\begin{figure}[H]
    \includegraphics[width=8cm]{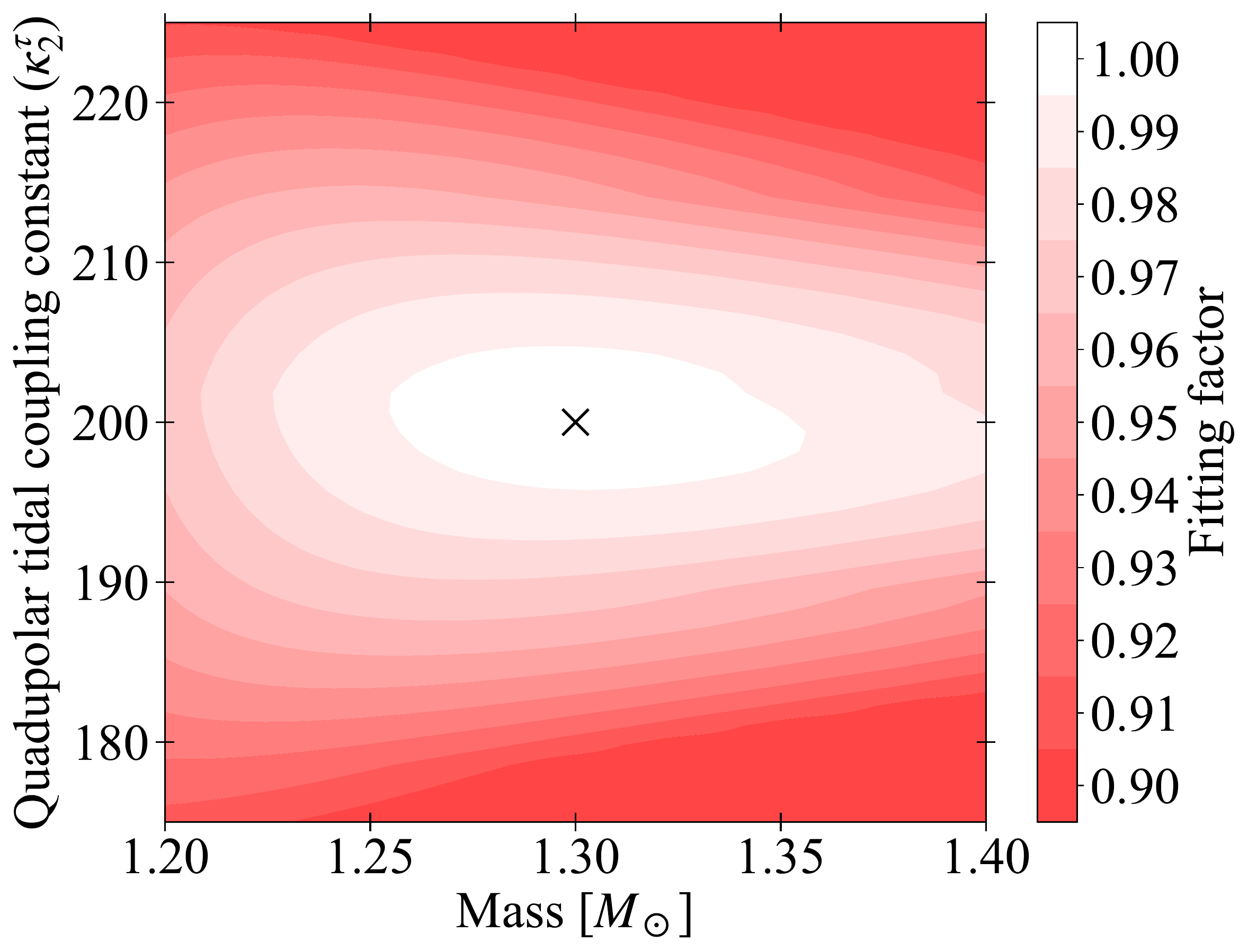}
        \caption{Fitting factor between spectra generated using our hierarchical model with 1.30\,$M_{\odot}$ and $\kappa_2^\tau=$200 (black cross), against a grid of mass and quadrupolar tidal deformability} values.
        \label{fig:GeneratedFittingFactor}
\end{figure}

        In Fig.~\ref{fig:GeneratedFittingFactor} we compare the fitting factor between a spectrum generated with $M=$~1.3\,$M_\odot$, $\kappa_2^\tau=$~200 against spectra generated at other parameter values. The location of the reference spectrum is shown with the black cross. This provides the first indication of whether this model could recover the mass and quadrupolar tidal deformability when trained on sufficient numerical-relativity simulations. \par
        The peak of the contour plot around the reference waveform shows that this model is selective, and may be used for parameter estimation and/or detection in the future. However, this is a task for future work and will require full Bayesian analysis with a noise implementation. This will ultimately allow for a comparison of the post-merger and inspiral estimate of the quadrupolar tidal deformability,  which may help determine whether a phase transition occurs in the neutron star equation of state~(e.g.~\cite{Bauswein2018,Most2018}). We leave the exploration of this as future work. If posteriors for the mass and tidal deformability are able to be determined, then it is a simple step to calculate the posteriors for the compactness, Eq. ~(\ref{eq:Ccalc}), and the radius of the neutron star.\par 
        In this letter, we use a hierarchical model to generate binary neutron star  post-merger spectra by training on spectra generated with numerical-relativity simulations. Our trained model allows us to generate spectra in ${\sim}10^{-4}$ seconds, which significantly reduces the computational effort required to populate a template of spectra. We obtain noise-weighted, amplitude-only fitting-factors across all tested spectra, with a mean of 0.95 and a standard deviation of 0.03, for sources simulated at a distance of 50 Mpc. This is better than post-merger fitting factors between different numerical-relativity codes,  which we find to be 0.76 and 0.85 for H4 and SLy, respectively.\par
        Training on the post-merger phase will allow fitting-factor comparisons with both the amplitude and phase information, as well as the generation of time-based waveforms. In addition, it will provide insight on the number of numerical-relativity simulations required to achieve a complete and accurate database. While obtaining information on the phase evolution is in principle possible, see, e.g.,~\cite{Bose2018}, this also requires a systematic investigation that goes well beyond the scope of this work. Without the phase information, a matched filter search is less sensitive, but it is still possible to design such a search using only the signal amplitudes.
        \par
        Results based on our trained model suggest that the model is selective and could potentially be used in parameter estimation of detected events. This will be confirmed in future work using a Bayesian framework. Parameter estimation of the post-merger spectra could set bounds on the post-merger quadrupolar tidal deformability, allowing comparison with the inspiral value. This could determine whether a phase change in the equation of state is present~\cite{Bauswein2018,Most2018}.\par
        To be valuable in search and parameter-estimation studies, our model must be extended to include individual values of mass, spins, compactness and quadrupolar tidal deformabilities for each progenitor.  These changes can be introduced given enough numerical-relativity simulations to cover the required ranges of mass and spin values. The placement of numerical-relativity simulations to enable this is a subject of future work. \par
        Our method may eventually provide an additional tool to aid in the detection of short-term post-merger neutron star remnants, supplementing the existing tools~\cite{Klimenko2016,Chatziioannou2017}.
\section{Acknowledgments}
        PDL is supported through Australian Research Council (ARC) Future Fellowship FT160100112 and ARC Discovery Project DP180103155, ARC is supported in part by the Australian Research Council through Discovery Grant DP160100637. We thank James Clark and Eric Thrane for their feedback.
\bibliography{FinalBib}
\end{document}